# On the Atmospheric Neutrino Anomaly
# and its Statistical Significance

G.L. Fogli[1] and E. Lisi[1,2]

[1]*Dipartimento di Fisica and Istituto Nazionale di Fisica Nucleare, Bari, Italy*
[2]*Institute for Advanced Study, Princeton, NJ 08540*

## Abstract

An analysis of the existing data on the atmospheric neutrino anomaly is presented, focused on the statistical significance that can be attributed to its experimental evidence. Our approach is alternative to the usual analyses in terms of the $\mu/e$ ratio of event rates. In fact, we perform a comparison between data and expectations, by *separating* the information on $e$-like and $\mu$-like events, with a careful estimate of the different errors and of their correlation effects. The results are shown both numerically and graphically, and disclose interesting aspects of the atmospheric neutrino anomaly, that the use of the $\mu/e$ ratio would partially hide, both in the sub-GeV and in the multi-GeV energy range.

## 1  Introduction

"Atmospheric neutrino anomaly" is usually referred to as the unexpected difference between measured and predicted muon/electron flavor composition of the atmospheric neutrino flux. Claimed as possible evidence of new physics beyond the Standard Model of electroweak interactions, it is generally interpreted in terms of neutrino oscillations.

First pointed out by the Kamiokande collaboration [1], the evidence for an anomaly in the sub-GeV energy range ($\langle E_\nu \rangle \lesssim 1$ GeV) has been reinforced in further exposures of the same Kamiokande detector [2, 3], and has also been confirmed by a similar (water-Cherenkov) underground experiment, IMB [4, 5]. More recently, the Kamiokande collaboration has reported that the flavor composition of a higher energy event sample (the so-called multi-GeV events) is also anomalous [3]. On the other hand, two of the iron-calorimeter experiments, Fréjus [6, 7] and NUSEX [8], did not find results in conflict with the expectations. The third, Soudan 2 [9], possibly does, although its data analysis is still preliminary.

Actually, a comparison of the experimental data with the expectations requires a reliable and precise calculation of the (anti)neutrino fluxes and their flavor composition. Conversely, as is well known, we observe a large spread among the different independent atmospheric $\nu$ flux calculations, hereafter referred to as: BGS (Barr, Gaisser and Stanev) [10], P (Perkins) [11], HKHM (Honda et al.) [12], LK (Lee and Koh) [13], BN (Bugaev and Naumov) [14], KM (Kawasaki and Mizuta) [15] [1]. This spread reflects essentially the large uncertainty in the overall normalization (of order 20–30%). The uncertainty, however, is reduced to a value as low as 5% when the $\mu/e$ flavor ratio is considered [17]. This cancellation of errors is the main motivation for the commonly used double ratio

$$R_{\mu/e} = (\mu/e)_{Data}/(\mu/e)_{MC} \qquad (1)$$

(see, e.g., the systematic review [18]) where $\mu$, $e$ represent, respectively, the number of $\mu$-like and $e$-like events in a given detector, as observed (Data) or simulated through a Monte Carlo (MC) numerical experiment.

The (often forgotten) drawback of the above double ratio is that, by construction, its error distribution is non-gaussian. A typical "anomalous" result, such as $R_{\mu/e} = 0.5 \pm 0.1$, represents thus an uncomplete summary of the data that could mislead the naïve reader, if not supplemented by the true error distribution. Such distribution should be given explicitly by the experimental collaborations, since it involves the knowledge of the errors *before* the ratio is taken. The use of two slightly asymmetric errors on $R_{\mu/e}$ as in Ref. [3], is only of limited help, and certainly is not sufficient to recover, for people other than the experimentalists themselves, the full information needed for correct statistical tests and phenomenological analyses.

Our point of view is that it is possible (and perhaps easier) to use exclusively gaussian distributions, and also to exploit fully the experimental and theoretical information, by separating the $e$ and $\mu$ flavor data. The modest price to pay is that the correlation of the different variables must be accounted for. More precisely, we will use and compare only those variables whose errors can be assumed to be normal (at least in the absence of any *a priori* contrary reason), that is: $\mu_{MC}$, $\mu_{Data}$, $e_{MC}$ and $e_{Data}$.

In the following sections this approach will be presented and developed systematically, in order to assess the significance of the atmospheric neutrino anomaly, as revealed (or not) in the various experiments performed so far. More precisely, in Section 2 we discuss in detail the experimental and simulated data, with a careful analysis of their errors and correlation effects. Section 3 is devoted to a comparative discussion of the results, with a specific example of a possible interpretation in terms of neutrino oscillations. In Section 4 we consider the very interesting analysis of the multi-GeV events, binned according to their direction. Finally, in Section 5 we draw our conclusions [2].

---

[1] All these neutrino flux calculations include the muon polarization effect, first pointed out by Volkova [16].

# 2    Analysis of $(\mu, e)_{Data}$ and $(\mu, e)_{MC}$ uncertainties

In this Section the gaussian errors affecting the doublet of variables $(\mu, e)_{Data}$ and $(\mu, e)_{MC}$ are estimated together with their correlations. We discuss first the largest, common, source of uncertainty in $(\mu, e)_{MC}$, induced by the spread in the theoretical calculation of the neutrino fluxes. Then, we estimate all the other relevant error sources for each specific experiment.

As far as neutrino fluxes are concerned, let us consider the predicted $(\mu, e)$ rates for the Kamiokande detector in the sub-GeV range, as calculated by using the BGS input fluxes (from this point forward, we conventionally rescale all rates to the central value of the BGS rates). Accordingly, the theoretical predictions, treated as a statistical population, can be conveniently represented by a bivariate gaussian distribution in $(\mu/\mu_{BGS}, e/e_{BGS})$ with the center at $(1, 1)$, errors $s_e$ and $s_\mu$ as large as 30% and correlation $\rho = 0.986$. The standard deviation ellipse[3] corresponding to such distribution is shown in Fig. 1. In the same figure we also show the alternative predictions (only the central values) coming from the other different input fluxes (P, HKHM, LK, BN, KM). In particular, the spread in the KM predictions (dots) is reminiscent of the flux variations obtained by the authors by varying a few input parameters [15]. Figures for other detectors (not shown) would be similar. In Fig. 1 we also draw some isolines of the double ratio $(\mu/e)/(\mu/e)_{BGS}$ (dashed), that will prove useful in the following.

Let us briefly comment some of our previous choices:

1) We choose to center the distribution on the BGS predictions because the BGS neutrino spectra are very well documented and detailed, and have also been used by all the experimental collaborations in at least one simulation, so that they are appropriate for a global comparison.

2) A $1\sigma$ uncertainty as large as 30% accounts conservatively for those flux calculations having the smallest normalization (LK and BN), and, at the same time, reduces much of the model-dependence implicit in the previous choice of BGS as reference fluxes. The correlation value $\rho = 0.986$ guarantees a residual $s_{\mu/e} = 5\%$ theoretical error on the $\mu/e$ ratio.

---

[2] This work does not cover the phenomenology of neutrino-induced upward-going muons, for which there are already extensive analyses [19, 20] as well as critical insights [21, 22].

[3] We use the notation of Ref. [23]. The standard deviation ellipse corresponds to the 39% C.L. in two variables; its projections onto the axes correspond to $\pm 1\sigma$ errors on $e$, $\mu$. The ellipse containing 68% (90%) probability would be similar, but rescaled linearly by a factor 1.51 (2.15).

**3)** In principle, the flux uncertainty could be substantially reduced [17] by using as an additional constraint the data on negative muon fluxes, as those reported in [24]. This constraint is used explicitly in Ref. [11] (which is in good agreement with BGS) and in a very recent and detailed flux calculation [25]. In particular, in the latter work the value $s_{\mu/e} = 5\%$ is confirmed, and the estimate $s_\mu = s_e = 20\%$ is defended. We will thus supplement our "default" case ($s_{flux} = 30\%$, used in all figures) with the less pessimistic error estimate $s_{flux} = 20\%$. In this case, the corresponding correlation is $\rho = 0.969$. The purely hypothetical case $s_\mu = s_e = 10\%$ will be also discussed.

A further comment on Fig. 1 is in order. A large cancellation of the theoretical flux uncertainties down to 5% can be considered established only for the ratio of the *total* rates, but is not guaranteed in small subsamples of the MC events, as, e.g., the five bins of the angular distribution of multi-GeV events (Ref. [3], Fig. 4), that will be studied in Sec. 4. Indeed, as far as the theoretical angular flux distributions shown in Ref. [15] (HKHM) and Ref. [26] (KM) can be compared, the bin-by-bin differences in $\nu_\mu/\nu_e$ appear to be larger than 5%. The same trend, i.e. larger uncertainties in each bin than in the integrated rate, characterizes the $\nu_\mu$ fluxes at the higher energies relevant for upward-going muon production [19, 27]. Lacking a detailed comparison of the different predicted neutrino angular distributions, it seems reasonable to assume an uncertainty of 10% on the $\mu/e$ ratio in each of the above five bins, this choice being more conservative than that performed in the Kamiokande analysis [3].

The above theoretical uncertainties are shared by all the experiments. Now we have to consider the remaining detector-dependent errors, which affect both $(\mu, e)_{Data}$ and $(\mu, e)_{MC}$ in each different experiment. Thus, we list in Table 1 the absolute (published) values of $(\mu, e)_{Data}$ and $(\mu, e)_{MC}$ for the various atmospheric neutrino experiments. For the reasons given before, all the simulated numbers correspond to BGS input fluxes. Also shown in Table 1 are the total and simulated exposures for each detector. Concerning the Kamiokande experiment, the reported data refer to both the sub-GeV and the multi-GeV energy range, the latter including fully contained (FC) and partially contained (PC) events [3]. The FC and PC samples are characterized by $\langle E_\nu \rangle \simeq 3$ GeV and $\langle E_\nu \rangle \simeq 9$ GeV, respectively. In the multi-GeV range, the quoted number of MC events corresponds, more precisely, to BGS fluxes supplemented with Volkova fluxes [28] for $E_\nu \gtrsim$ few GeV ("flux B" of Ref. [3]). The additional information provided by the zenith-angle distribution of multi-GeV data will be examined separately in Sec. 4. For the IMB detector [4, 5], only the contained event sample is considered. The data for Fréjus in Table 1 include the fully contained (FC) events, and the total sample (ALL), as reported in [6, 7]. Concerning the NUSEX experiment, the predictions apparently refer [8] to the BGS fluxes *without* muon polarization [29]. In view of the very large errors in this experiment, we have not attempted any correction. Our source for the NUSEX simulated exposure is Ref. [30]. In the last row of Table 1 we consider the preliminary Soudan 2 data [9]: in this case, the number of observed events is not an integer, due to a "shield inefficiency correction" [9].

We now present our analysis of the uncertainties. Although only published data are used (unless otherwise noticed), our method differs from the usual approaches, in that we are interested in the errors affecting the *separate* flavors $\mu$ and $e$, and not only the flavor ratio $R_{\mu/e}$. This requires that the error correlations $\rho_{\mu e}$ are to be taken into account, as done previously for the flux uncertainties. The reader is referred to Table 2, where we have collected the actual values of the individual sources of errors and their combinations.

In the discussion of Table 2, let us first consider the data errors. Of course, all data samples are affected by statistical errors (with $\rho_{\mu e} = 0$), obtainable from the values of $\mu_{Data}$ and $e_{Data}$ of Table 1. All the experiments also consider the possibility of flavor misidentification (mis-id), for which we have taken the published values, with full anti-correlation: $\rho_{\mu e} = -1$. Kamiokande lists several additional sources of data errors: multiring event separation, vertex fit, absolute energy calibration and non-neutrino background. For sub-GeV data, they add up to $s_{\mu/e} \simeq 2.7\%$. In the absence of any other information, in Table 2 their correlation is disregarded and they are assumed as equally shared between $\mu$ and $e$: $s_e = s_\mu = 2\%$. Analogously, for multi-GeV data it is $s_{\mu/e} \simeq 5.4\%$ [3], so that $s_e = s_\mu = 3.8\%$. Concerning the Fréjus experiment, the trigger efficiency uncertainties [6] should also be considered. They can be disregarded for $\mu$-like events, but are sizeable for $e$-like events: $s_e = 10\%$. No other sources of data errors are quantitatively discussed in the NUSEX paper [8] and in the Soudan 2 report [9]. Finally, the "Total" errors in Table 2 are obtained by summing in quadrature all the $2 \times 2$ error matrices associated to the values of $s_\mu$, $s_e$ and $\rho_{\mu e}$ considered so far, for each experiment separately.

Let us now consider the MC errors, i.e. those affecting the simulation of event production in each detector. The largest contribution, provided by the neutrino flux uncertainties, has already been discussed at length, and is not reported in Table 2. For all the simulated samples, the statistical errors must be taken into account, according to a binomial distribution of $\mu$ and $e$ events. The relevant input can be taken from Table 1. An explicit check of our estimate is possible for Kamiokande, the quoted statistical MC errors being $s_{\mu/e} = 3.6\%$ [26] and $s_{\mu/e} = 6\%$ [3] for the sub-GeV and FC+PC multi-GeV cases, respectively. Our estimates in Table 2 imply that $s_{\mu/e} = 3.4\%$ and $s_{\mu/e} = 5.6\%$ respectively (the agreement would be even better by using HKHM instead of BGS fluxes). Concerning the MC errors related to the neutrino interaction in the detectors, cross section and nuclear model uncertainties have been treated differently by the various collaborations. For Kamiokande sub-GeV, charged current (CC) cross section errors are estimated to amount to $\sim 10\%$ for each flavor [31], reduced to $3\%$ in the $\mu/e$ ratio [26], $\rho_{\mu e}$ being determined by this cancellation. Neutral current (NC) cross section uncertainties affect mainly $e_{MC}$ through $\pi^0$ contamination in the Kamiokande sample: $s_e \simeq 2\%$, $s_\mu \simeq 0$. These numbers are slightly different for the multi-GeV case: $s_{\mu/e}(\text{CC}) = 2\%$ and $s_e(\text{NC}) = 3\%$ [3]. The IMB collaboration estimates nuclear and cross section uncertainties more conservatively than Kamiokande, although it is difficult to extract definite error values from the published papers [4, 5]. The single largest effect in the IMB simulation is induced by varying by $\pm 20\%$ the axial mass parameter, leading to $s_\mu \simeq s_e \simeq 20\%$ and $s_{\mu/e} \simeq 10\%$ [32]. Fermi gas model uncertainties are estimated to increase $s_{\mu/e}$ up to $\sim 14\%$. The final values of "nucl.+cross" in Table 2 for IMB correspond to the choice $(s_\mu, s_e, s_{\mu/e}) = (20, 20, 14)$. For the Fréjus experiment, nuclear and cross-section uncertainties amount to $s_{\mu/e} = 6\%$ [7]. Values of $s_\mu$ and $s_e$ are not published, however taking $s_\mu \simeq s_e \simeq 10\%$ (similar to the Kamiokande case) is not unreasonable, and leads to the values in Table 2. Concerning the NUSEX and Soudan 2 experiments, in the absence of detailed published information we have assumed nuclear uncertainties as large as for Fréjus, a choice which hardly leads to overestimate the errors. Finally, the "Total" MC errors reported in Table 2 are obtained by adding in quadrature the previous errors *and* the flux uncertainties (only the case $s_{flux} = 30\%$ is explicitly reported).

# 3   Synopsis of the Results

We have estimated, as carefully as we could, the errors affecting $(\mu, e)_{Data}$ and $(\mu, e)_{MC}$ for each experiment. There are no *a priori* reasons against the assignment of gaussian distributions to these errors, except perhaps for the fluctuations in the (small) number of events observed by NUSEX, which should be described more properly by a Poisson distribution. However, data and simulations do agree in NUSEX, so that the tail of the statistical error distribution is not probed, and the Poisson distribution can be well approximated by a gaussian within $\pm 1\sigma$.

Thus, an unbiased $\chi^2$ can be defined for each experiment:

$$\chi^2 = \sum_{\alpha, \beta = \mu, e} \Delta_\alpha \cdot \left( \sigma^2_{Data} + \sigma^2_{MC} \right)^{-1}_{\alpha\beta} \cdot \Delta_\beta \quad , \qquad (2)$$

where $\sigma^2$ are the $2 \times 2$ total (squared) error matrices, and $\Delta$ is the vector of the Data$-$MC differences: $\Delta = (\mu_{Data} - \mu_{MC}, \, e_{Data} - e_{MC})$.

The list of $\chi^2$ values and corresponding C.L. (d.o.f. = 2) for the experiments examined in the previous Section is given in Table 3. The second and third column refer to our "default" case ($s_{flux} = 30\%$), the next two columns to the estimate $s_{flux} = 20\%$ reported in [25] and the last two columns to the purely hypothetical case $s_{flux} = 10\%$. In any given column, the "hierarchy" of experimental evidences for the atmospheric neutrino anomaly is rather evident and does not need any comment. The comparison between the columns at $s_{flux} = 30\%$ and $20\%$ shows that the results are remarkably stable with respect to the normalization error, especially for the experiments that do show an anomaly, the reason being that, in this case, the greatest contribution to the $\chi^2$ is related to the uncertainties *orthogonal* to $s_{flux}$. A further hypothetical reduction down to $10\%$ breaks this stability, increasing a few $\chi^2$ values much more than others; in this situation, however, also the specific choice of a "reference" flux (BGS in our case) starts becoming crucial for the results.

The above results can be better understood by means of Fig. 2, where the standard error ellipses (39% C.L.) of data (white) and MC (gray) are displayed for each experiment, together with the corresponding iso-$R_{\mu/e}$ lines (dashed). All MC ellipse include a 30% flux error. The IMB and Soudan 2 MC ellipses are the largest ones, as a result of more conservative nuclear error estimates (IMB) or small simulated statistics (Soudan 2). The narrowness of the NUSEX Monte Carlo ellipse only reflects our ignorance of the corresponding systematics.

Before commenting on the single experiments shown in Fig. 2, it should be noted that any change in the overall normalization of the fluxes (e.g., any choice of reference fluxes different from BGS) would only have the effect of shifting the MC gray ellipses up or down along their major axis. Moreover, such shifts are bound to be approximately equal for all those experiments which are sensitive to the *same range* of the neutrino energy spectrum. In particular, this property holds for NUSEX, Fréjus (FC), Kamiokande sub-GeV, IMB, and possibly Soudan 2: all of them, in fact, observe contained events and are sensitive to $0.2 \lesssim E_\nu \lesssim 2$ GeV.

Concerning the iron-calorimeter experiments (Fig 2a,b,c), the close agreement between the NUSEX and Frejus results on $\mu$ and $e$ separately is remarkable, being more informative than simply the agreement of the $R_{\mu/e}$ values (dashed lines). Both experiments seem to favor fluxes with low normalization. The preliminary Soudan 2 data are slightly far from the expectations. Fig. 2c provides the additional information that Soudan 2 data favor fluxes with "central normalization". Assuming a theoretical error lower than 30% in the MC ellipses, the different indications provided by the three iron detectors would be correspondingly exacerbated.

Concerning water-Cherenkov experiments (Fig. 2d,e,f), it is the very good agreement between the data of the two high-statistics experiments IMB and Kamiokande sub-GeV (white ellipses), and their common disagreement with the MC simulations (gray ellipses), which provides the well-known evidence for an anomaly in the sub-GeV range. It is interesting to observe, however, that the standard deviation error ellipses of the IMB, Kamiokande sub-GeV, Frejus FC, NUSEX and Soudan 2 data are mutually compatible. The relative position of these five data ellipses would not be spoiled by choosing a reference flux different from BGS: as said, that would simply correspond to shift the five MC ellipses by one and the same amount. Thus, this additional degree of freedom cannot bring to a closer agreement the indications coming from Fréjus and NUSEX on the one hand, and Kamiokande sub-GeV and IMB on the other hand.

Multi-GeV data (Fig. 2f) show an additional feature. If we limit our attention only to the ratio $R_{\mu/e}$, the results from the sub- and multi-GeV samples of Kamiokande agree rather well (slanted lines of Figs. 2e,f). But this is not the case for the separate $\mu$ and $e$ flavours, as it clearly emerges from the different position of the data ellipses. This does not imply an inconsistency between the two data sets: simply, it seems to indicate that the ratios $\mu/\mu_{BGS}$ and $e/e_{BGS}$ increase with energy, while keeping the double ratio $R_{\mu/e}$ approximately constant. We note that part of this effect could be explained, for instance, by assuming a corresponding hypothetical decrease in the slope of the theoretical neutrino energy spectra. With regard to this, future re-calculations of atmospheric fluxes can usefully address the problem of estimating the allowed range of such *shape* variations.

The more abundant information coming from the comparison of sub- *and* multi-GeV data (not even including the directional information of the multi-GeV data) can make the usual attempts to explain the atmospheric neutrino anomaly less feasible. As an example, we show in Fig. 3 what happens by assuming pure $\nu_\mu \to \nu_\tau$ oscillations with a high value of $\Delta m^2$, so that $P(\nu_\mu \to \nu_\tau) \simeq \frac{1}{2} \sin^2 2\theta$, *independently* on the energy (the $e$ flavor is not affected at all). All error ellipses are considered in the same plot, whereas, in order to avoid confusion, only a representative MC ellipse is shown, both in the standard case (dashed) and assuming oscillations (gray). The choice $\sin^2 2\theta = 0.66$ is seen to bring the theoretical predictions in agreement with the data coming from the different experiment. However, the sub- and multi-GeV Kamiokande data clearly favor *different* normalization of the MC rates. With theoretical errors much smaller than 30%, $\nu_\mu \to \nu_\tau$ oscillations with high $\Delta m^2$ would thus be unable to fit both sub- and multi-GeV data at the same time: this could be a tentative indication in favor of additional effects, able to renormalize the rates in an energy-dependent way.

Apart from the above example, and another sketchy analysis of $\nu_\mu \leftrightarrow \nu_\tau$ mixing in the next Section, we do not further investigate in this paper neutrino oscillations, as an exhaustive analysis of them requires, in our opinion, a separate work [33].

# 4    Analysing Multi-GeV Binned Data

So far we have analyzed only total rates, and not the full information contained in convenient subsamples, as the histograms reporting energy and/or angle distributions. It is difficult to perform accurate "binned analyses," for at least two reasons. On the one hand, bin-by-bin experimental systematics are usually not published, and cannot even be guessed. One the other hand, theoretical uncertainties are less under control, since their dependence on the neutrino energy and/or direction is largely unknown.

Nevertheless, a few interesting insights can be gained by looking at a specific example: the angular distribution of multi-GeV data in Kamiokande [3]. This is a particularly important case, since the evidence for an anomalous angular distribution would strengthen the neutrino oscillation hypothesis. We use the information contained in Figs. 3a and 3b of Ref. [3], where the data were divided into five bins equally spaced in the cosine of the zenith angle. The first (fifth) bin correspond to upward (downward) neutrinos. In each bin, we compare $(\mu, e)_{Data}$ with $(\mu, e)_{MC}$ as it has been done before for the total rates, the only exception being that now we allow $(s_{\mu/e})_{flux} = 10\%$, as discussed in Sec. 2.

The results are shown in the upper part of Fig. 4, reported as the "no-oscillation case," with the uncertainties drawn as gray (MC) and white (Data) ellipses. Data errors are assumed gaussian, although a Poisson distribution would be more appropriate to describe small data samples. This choice, however, would imply making the variables $(\mu, e)_{Data}$ discrete, thus preventing any simple graphical representation. Moreover, it can easily be checked that for the integer values of $(\mu, e)_{Data}$ of interest in Fig. 4, the difference between poissonian and gaussian $\pm 1\sigma$ intervals is small, and in any case irrelevant for the following discussion. Coming back then to this "no-oscillation case," we see that the agreement between data and MC simulation, quite bad in the 1st bin, increases in the next four bins.

A better fit can be obtained, however, by assuming a neutrino oscillation scenario. Again, the simple case of pure $\nu_\mu \rightarrow \nu_\tau$ oscillations is considered. In particular, the medium strip of Fig. 4 displays the results obtained for maximal mixing ($\sin^2 2\theta = 1$) and large $\Delta m^2$ ($\Delta m^2 > 10^{-1}$ eV$^2$), so that $\mu_{MC} \rightarrow \frac{1}{2}\mu_{MC}$. In this case, the agreement between data and MC simulation is lost in the 5th bin, but improves in the first 4 bins.

An even better fit can be obtained by lowering $\Delta m^2$. In particular, the lower part of Fig. 4 shows the analysis for the best fit values of the mass/mixing parameters, as taken from Ref. [3]. Now the agreement between data and MC simulation is improved with respect to both the previous cases. It is instructive to observe that the simple comparison of $\mu/e$ ratios, corresponding to the slopes of the slanted lines, would have hidden the persistent discrepancy between data and MC simulation in the first bin.

We do not attach any definite C.L. to the three scenarios shown in Fig. 4, because of the aforementioned ignorance of potentially important bin-by-bin correlations, and leave the reader to judge the significance of the angular anomaly and its explanation in terms of oscillations. As far as our opinion is concerned, we quite reasonably conclude that the evidence for the third (oscillation) scenario, although significant, is not really striking. In particular, it should be noted that the 1st and the 5th bin, which play an important role in the fit, contain the smallest number of events, the angular $\mu$ and $e$ distributions being both peaked at the central bin [3].

It must be noted that the information used by the Kamiokande collaboration for statistical tests is larger than the angular distribution alone, and includes a $8 \times 5$ (energy$\times$angle) histogram of MC and Data events. Unfortunately, only the energy and angle projections are published [3]. We hope that in future publications this additional, and potentially very important, spectral information will be fully reported.

## 5 Conclusions

The atmospheric neutrino data coming from a large part of the experiments performed so far show an interesting pattern of deviations with respect to the theoretical predictions, that is usually summarized in the double ratio $R_{\mu/e}$. This requires, however, an analysis of the non-gaussian distribution of the ratio uncertainties.

An alternative approach has been proposed in this paper, which allows to use only gaussian distributions, provided that the information on the the $\mu$ and $e$ flavors are separated, and their correlation effects are properly taken into account.

Accordingly, we have performed for each experiment a careful analysis of the correlated uncertainties affecting the e-like and mu-like observed and simulated event rates. This enabled us to assess quantitatively the statistical significance of the atmospheric neutrino anomaly in each experiment, and to study the sensitivity of the results to the flux uncertainties.

The anomalous angular distribution of the Kamiokande multi-GeV data has been also analyzed, separately, with the same methodology. It is shown graphically that neutrino oscillations can bring the expectations in closer agreement with the data, although more data are needed to derive compelling indications.

In conclusion, we have shown in detail an unconventional way of looking at the atmospheric neutrino anomaly, in which the slight complication of taking into account non-zero error correlations is more than compensated by the benefit of having clear graphical and numerical results.

# Acknowledgements

Useful conversations with D. Casper and T. Stanev are gratefully acknowledged. The work of E.L. is supported in part by a post-doc INFN fellowship and by funds of the Institute for Advanced Study. This research was in part performed under the auspices of the Theoretical Astroparticle Network, under contract N. CHRX-CT93-0120 of the Direction General 12 of the E.E.C.

# References

[1] Kamiokande Collaboration, K.S. Hirata et al., Phys. Lett. **B205** (1988) 416.

[2] Kamiokande Collaboration, K.S. Hirata et al., Phys. Lett. **B280** (1992) 146.

[3] Kamiokande Collaboration, Y. Fukuda et al., Phys. Lett. **B335** (1994) 237.

[4] IMB Collaboration, D. Casper et al., Phys. Rev. Lett. **66** (1991) 2561.

[5] IMB Collaboration, R. Becker-Szendy et al., Phys. Rev. **D46** (1992) 3720.

[6] Fréjus Collaboration, Ch. Berger et al., Phys. Lett. **B227** (1989) 489.

[7] Fréjus Collaboration, Ch. Berger et al., Phys. Lett. **B245** (1990) 305.

[8] NUSEX Collaboration, M. Aglietta et al.. Europhys. Lett. **8** (1989) 611.

[9] Soudan 2 Collaboration, M. Goodman et al., Nucl. Phys. **B** (Proc. Suppl.) **38** (1995) 337.

[10] G. Barr, T.K. Gaisser and T. Stanev, Phys. Rev. **D39** (1989) 3532.

[11] D.H. Perkins, Astropart. Phys. **2** (1994) 249.

[12] M. Honda, K. Kasahara, K. Hidaka and S. Midorikawa, Phys. Lett. **B248** (1990) 193.

[13] H. Lee and Y.S. Koh, Nuovo Cimento **105B** (1990) 883.

[14] E.V. Bugaev and V.A. Naumov, Phys. Lett. **B232** (1989) 391.

[15] M. Kawasaki and S. Mizuta, Phys. Rev. **D43** (1991) 2900.

[16] L.V. Volkova, in *Cosmic Gamma Rays, Neutrinos and Related Astrophysics*, Proceedings of the NATO Adv. Study Inst., ed. by M.M. Shapiro and J.P. Wefel, NATO ASI Series B: Physics Vol. 270 (Plenum, New York, 1989), p. 139.

[17] T.K. Gaisser, Nucl. Phys. **B** (Proc. Suppl.) **35** (1994) 209.

[18] E.W. Beier et al., Phys. Lett. **B283** (1992) 446.

[19] W. Frati, T.K. Gaisser and A.K. Mann, Phys. Rev. **D48** (1993) 1140.

[20] E. Akhmedov, P. Lipari and M. Lusignoli, Phys. Lett. **B300** (1993) 128.

[21] D.H. Perkins, Nucl. Phys. **B399** (1993) 3.

[22] P. Lipari, M. Lusignoli and F. Sartogo, Phys. Rev. Lett. **74** (1995) 4384.

[23] Particle Data Group, *Review of Particle Properties*, Phys. Rev. **D50** (1994) part I.

[24] M. Circella et al., in the *Proceedings of the 23rd Int. Cosmic Ray Conference, Calgary, 1993*, ed. by D.A. Leahy et al. (Univ. of Calgary, Canada 1993) Vol. 4, p. 503; see also R. Bellotti et al., Lab. Naz. Gran Sasso preprint LNGS 95/02, submitted to J. Geophis. Res.

[25] M. Honda, T. Kajita, K. Kasahara and S. Midorikawa, ICRR Report 336-95-2, hep-ph/9503439 (unpublished).

[26] T. Kajita, Lecture presented at the *5th BCSPIN Summer School in Physics* (Kathmandu, Nepal, 1994), ICRR Report 332-94-27.

[27] T. Stanev, private communication.

[28] L.V. Volkova, Yad. Fiz. **31** (1980) 1510 [Sov. J. Nucl. Phys. **31** (1980) 784].

[29] T.K. Gaisser, T. Stanev and G. Barr, Phys. Rev. **D38** (1988) 85.

[30] C. Broggini, private communication.

[31] Kamiokande Collaboration, Y. Totsuka et al., Nucl. Phys. **B** (Proc. Suppl.) **31** (1993) 428.

[32] D. Casper, private communication.

[33] G.L. Fogli and E. Lisi, in preparation.

# Figure Captions

**Fig. 1:**  Comparison between the different theoretical predictions of the rates of $\mu$-like and $e$-like events (in the sub-GeV range) for the Kamiokande detector, normalized to the predictions of the reference neutrino fluxes of Ref. [10] (BGS fluxes). The standard deviation ellipse (39% C.L.) assumed to represent the maximum (30%) theoretical dispersion is also shown. See the text for details.

**Fig. 2:**  Comparison between Monte Carlo simulations (gray ellipses) and experimental data (white ellipses) for the following experiments: NUSEX, Fréjus (ALL events and only fully contained, FC, events), Soudan 2 (preliminary), IMB, Kamiokande sub-GeV, Kamiokande multi-GeV (fully contained, FC, and partially contained, PC). Iso-lines of the double ratio $R_{\mu/e}$ are also shown (dashed).

**Fig. 3:**  Simultaneous comparison of the data with the theoretical (MC) predictions, including an example of neutrino oscillation effects. Dashed ellipse: MC, without neutrino oscillations. Gray ellipse: MC, with $\nu_\mu \to \nu_\tau$ oscillations at large $\Delta m^2$ and $\sin^2 2\theta = 0.66$.

**Fig. 4:**  Comparison of multi-GeV Kamiokande data (white ellipses) and simulations (gray ellipses) in each of the five zenith-angle bins. Upper strip: no neutrino oscillations. Medium strip: $\nu_\mu \to \nu_\tau$ oscillations with maximal mixing and large $\Delta m^2$. Lower strip: $\nu_\mu \to \nu_\tau$ oscillations with best fit mass/mixing parameters.

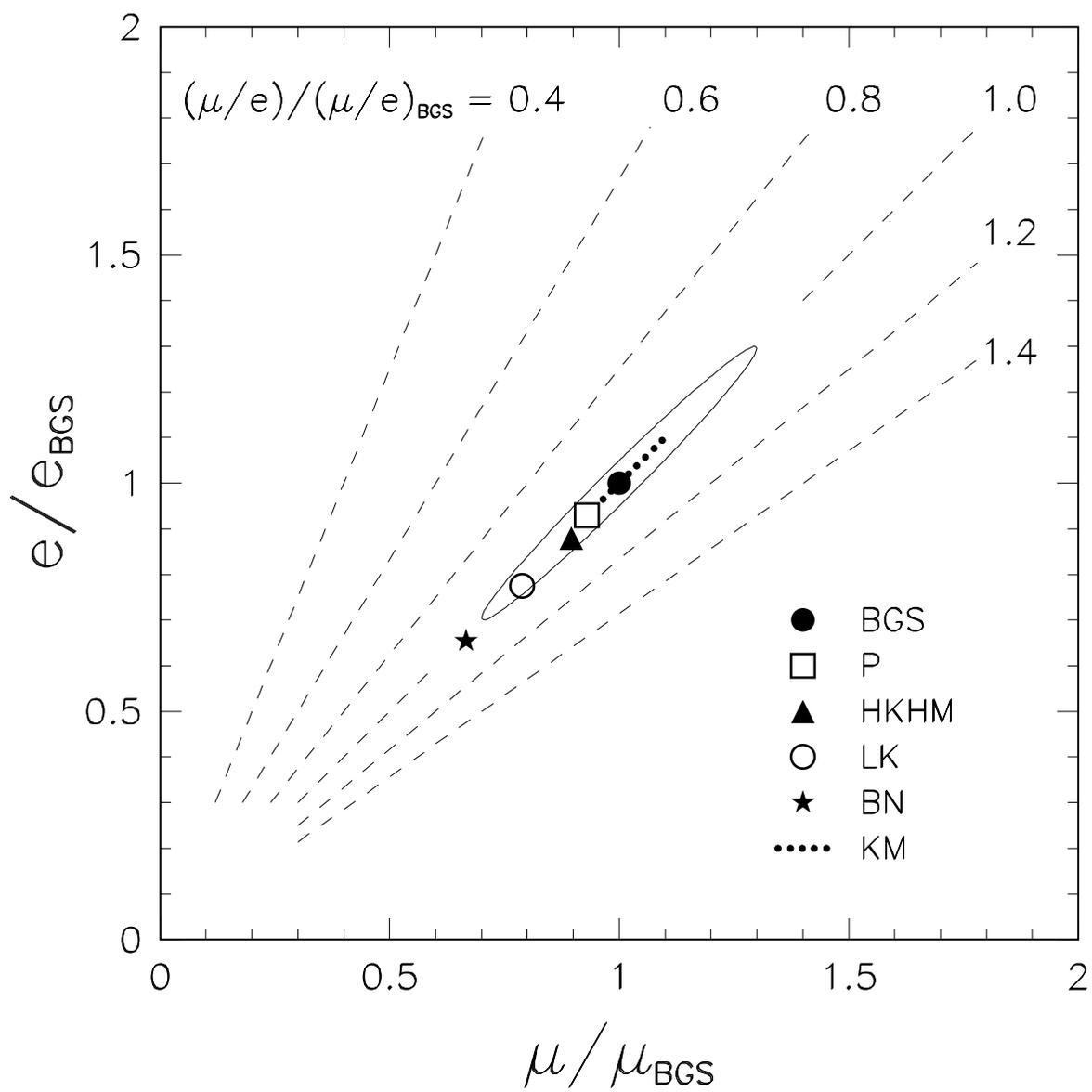

Fig. 1

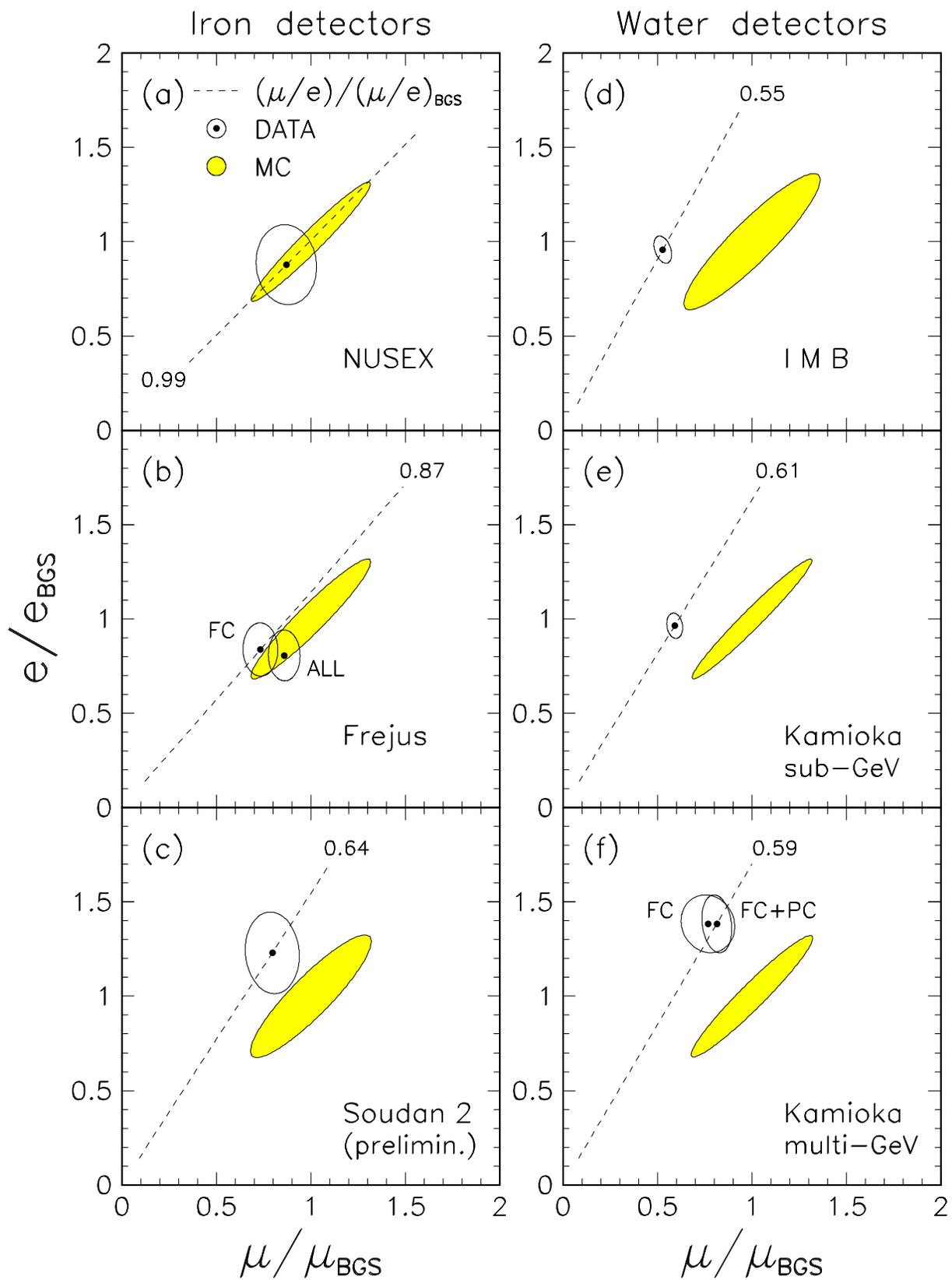

Iron detectors

Water detectors

(a) --- $(\mu/e)/(\mu/e)_{BGS}$
⊙ DATA
🟡 MC

NUSEX

0.99

0.55

I M B

$e/e_{BGS}$

(b)

FC

ALL

Frejus

0.87

(e)

0.61

Kamioka
sub-GeV

(c)

Soudan 2
(prelimin.)

0.64

(f)

FC    FC+PC

0.59

Kamioka
multi-GeV

$\mu/\mu_{BGS}$

Fig. 2

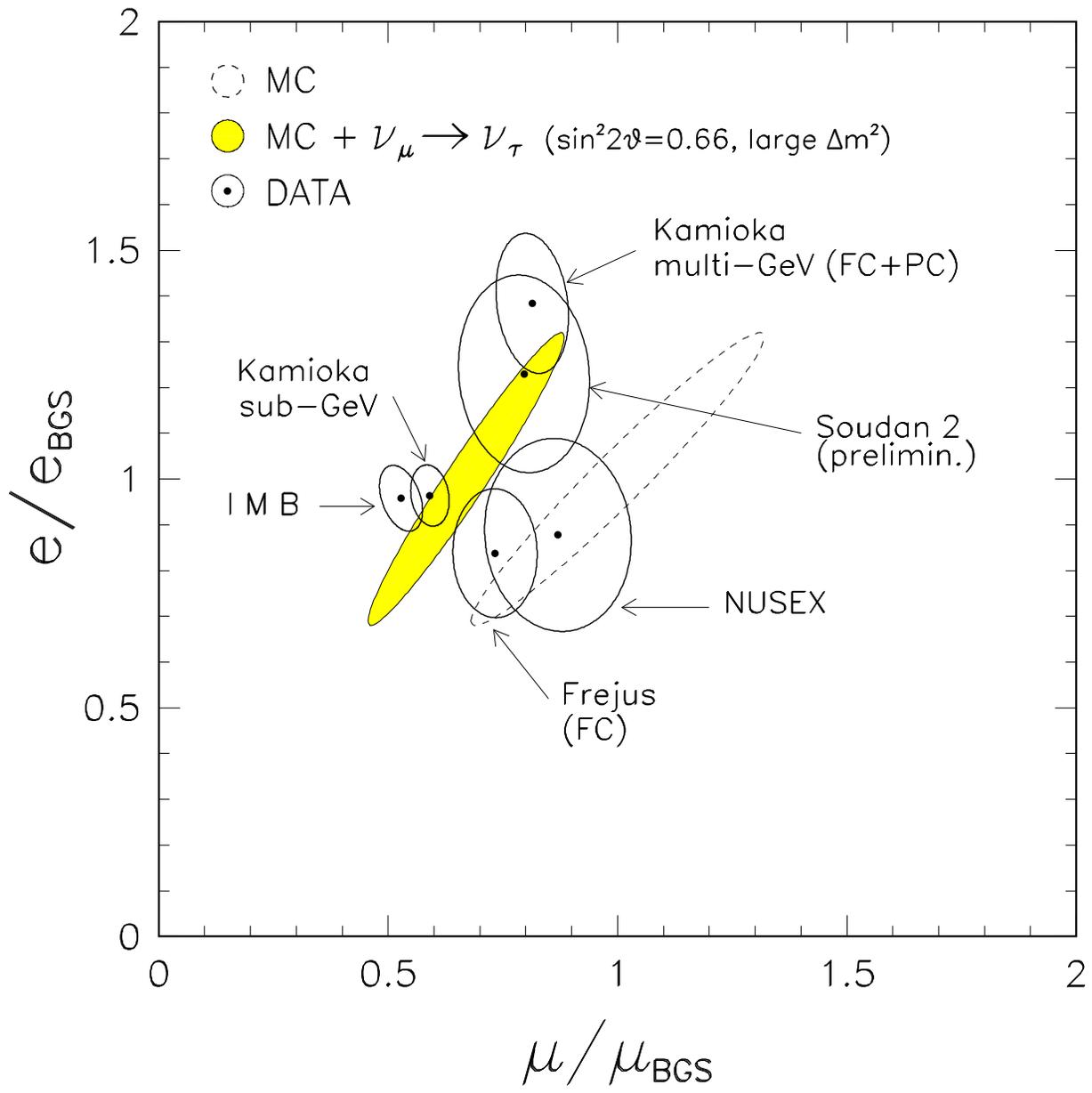

Kamioka multi-GeV data

Fig. 4

Table 1: Number of $\mu$-like and $e$-like events, as observed in each experiment (Data) or simulated with the reference fluxes of Ref. [10] (BGS). Event numbers refer to the total exposure of the detector (the exposure used in the MC simulation is also given).

| Experiment | Refs. | Total exposure (kt·yr) | Simulated exposure (kt·yr) | $\mu$-like events Data | $\mu$-like events BGS | $e$-like events Data | $e$-like events BGS |
|---|---|---|---|---|---|---|---|
| Kamiokande sub-GeV | [3] | 7.7 | 43 | 234 | 396.0 | 248 | 257.2 |
| Kamiokande multi-GeV (FC) | [3] | 8.2 | 51 | 31 | 40.4 | 98 | 70.8 |
| Kamiokande multi-GeV (FC+PC) | [3] | 8.2+6.0 | 51+40 | 135 | 165.8 | 98 | 70.8 |
| IMB | [4],[5] | 7.7 | 33 | 182 | 344.5 | 325 | 339.4 |
| Fréjus (FC) | [6],[7] | 1.56 | 10 | 66 | 90.0 | 56 | 66.8 |
| Fréjus (ALL) | [6],[7] | 1.56 | 10 | 108 | 125.8 | 57 | 70.6 |
| NUSEX | [8] | 0.74 | ∼15 | 32 | 36.8 | 18 | 20.5 |
| Soudan 2 (preliminary) | [9] | 1.01 | 3.72 | 33.5 | 42.1 | 35.3 | 28.7 |

Table 2: Individual and combined errors $s_\mu$ and $s_e$ (%), and their correlation $\rho_{\mu e}$, for both observed (Data) and simulated (MC) samples of $\mu$-like and $e$-like events. Total MC errors implicitly include neutrino flux uncertainties, characterized by: $(s_\mu, s_e, \rho_{\mu e}) = (30, 30, 0.986)$.

| Experiment | Data errors | | | | MC errors | | | |
|---|---|---|---|---|---|---|---|---|
| | Source | $s_\mu$ | $s_e$ | $\rho_{\mu e}$ | Source | $s_\mu$ | $s_e$ | $\rho_{\mu e}$ |
| Kamiokande sub-GeV | statistics | 6.5 | 6.4 | 0.000 | statistics | 1.3 | 2.1 | −1.000 |
| | mis-id. | 2.0 | 2.0 | −1.000 | CC cross sec. | 10.0 | 10.0 | 0.955 |
| | various | 2.0 | 2.0 | 0.000 | NC cross sec. | 0.0 | 2.0 | 0.000 |
| | Total | 7.1 | 7.0 | −0.081 | Total | 31.7 | 31.8 | 0.975 |
| Kamiokande multi-GeV (FC, FC+PC) | stat.(FC) | 18.0 | 10.1 | 0.000 | stat.(FC) | 5.0 | 2.9 | −1.000 |
| | stat.(FC+PC) | 8.6 | 10.1 | 0.000 | stat.(FC+PC) | 1.6 | 4.0 | −1.000 |
| | mis-id. | 2.0 | 2.0 | −1.000 | CC cross sec. | 10.0 | 10.0 | 0.980 |
| | various | 3.8 | 3.8 | 0.000 | NC cross sec. | 0.0 | 3.0 | 0.000 |
| | Tot.(FC) | 18.5 | 11.0 | −0.020 | Tot.(FC) | 32.0 | 31.9 | 0.951 |
| | Tot.(FC+PC) | 9.6 | 11.0 | −0.038 | Tot.(FC+PC) | 31.7 | 32.0 | 0.966 |
| IMB | statistics | 7.4 | 5.5 | 0.000 | statistics | 1.8 | 1.9 | −1.000 |
| | mis-id. | 5.0 | 5.0 | −1.000 | nucl.+cross | 20.0 | 20.0 | 0.755 |
| | Total | 8.9 | 7.5 | −0.374 | Total | 36.1 | 36.1 | 0.910 |
| Fréjus (FC, ALL) | stat.(FC) | 12.3 | 13.4 | 0.000 | stat.(FC) | 2.7 | 3.7 | −1.000 |
| | stat.(ALL) | 9.6 | 13.2 | 0.000 | stat.(ALL) | 2.1 | 3.8 | −1.000 |
| | mis-id. | 2.0 | 2.0 | −1.000 | nucl.+cross | 10.0 | 10.0 | 0.820 |
| | trigger eff. | 0.0 | 10.0 | 0.000 | | | | |
| | Tot.(FC) | 12.5 | 16.8 | −0.019 | Tot.(FC) | 31.7 | 31.8 | 0.950 |
| | Tot.(ALL) | 9.8 | 16.7 | −0.024 | Tot.(ALL) | 31.7 | 31.9 | 0.953 |
| NUSEX | statistics | 17.7 | 23.6 | 0.000 | statistics | 2.2 | 4.0 | −1.000 |
| | mis-id. | 5.0 | 5.0 | −1.000 | nucl.+cross | 10.0 | 10.0 | 0.820 |
| | Total | 18.4 | 24.1 | −0.057 | Total | 31.7 | 31.8 | 0.951 |
| Soudan 2 (prelimin.) | statistics | 17.3 | 16.8 | 0.000 | statistics | 5.0 | 7.5 | −1.000 |
| | mis-id. | 5.0 | 5.0 | −1.000 | nucl.+cross | 10.0 | 10.0 | 0.820 |
| | Total | 18.0 | 17.6 | −0.079 | Total | 32.0 | 32.5 | 0.896 |

Table 3: Values of $\chi^2$, and corresponding percentage C.L. (d.o.f. = 2) of the atmospheric neutrino anomaly hypothesis, for the experiments analyzed in Sec. 2. Flux errors as large as 30%, 20% and 10% are considered.

| Experiment | $s_{flux} = 30\%$ | | $s_{flux} = 20\%$ | | $s_{flux} = 10\%$ | |
|---|---|---|---|---|---|---|
| | $\chi^2$ | C.L. | $\chi^2$ | C.L. | $\chi^2$ | C.L. |
| Kamioka sub-GeV | 12.7 | 99.8 | 13.5 | 99.9 | 15.7 | 99.9 |
| Kamioka multi-GeV (FC) | 7.08 | 97.1 | 7.11 | 97.1 | 7.17 | 97.2 |
| Kamioka multi-GeV (FC+PC) | 8.79 | 98.8 | 8.81 | 98.8 | 8.86 | 98.8 |
| IMB | 6.13 | 95.3 | 6.53 | 96.2 | 14.3 | 99.9 |
| Fréjus (FC) | 0.79 | 32.6 | 1.26 | 46.7 | 2.45 | 70.6 |
| Fréjus (ALL) | 0.32 | 14.8 | 0.54 | 23.6 | 1.13 | 43.2 |
| NUSEX | 0.14 | 6.76 | 0.25 | 11.8 | 0.48 | 21.3 |
| Soudan 2 (preliminary) | 2.04 | 63.9 | 2.06 | 64.3 | 2.11 | 65.2 |